\documentclass[aerospace,article,accept,pdftex,moreauthors]{Definitions/mdpi} 

\usepackage{subcaption}

\firstpage{1} 
\pubvolume{10}
\issuenum{11}
\articlenumber{960}
\pubyear{2023}
\copyrightyear{2023}
\datereceived{28 September 2023} 
\daterevised{27 October 2023} 
\dateaccepted{30 October 2023} 
\datepublished{15 November 2023} 
\hreflink{https://doi.org/10.3390/{\linebreak}aerospace10110960} 

\usepackage{xspace, color}
\newcommand\superbit{\textsc{superBIT}\xspace}
\newcommand\drss{\textsc{drs}s\xspace}
\newcommand\drs{\textsc{drs}\xspace}
\newcommand\drsi{\textsc{drs1}\xspace}
\newcommand\drsii{\textsc{drs2}\xspace}
\newcommand\drsiii{\textsc{drs3}\xspace}
\newcommand\drsiv{\textsc{drs4}\xspace}
\newcommand\drsv{\textsc{drs5}\xspace}

\newcommand\nasa{\textsc{nasa}\xspace}
\newcommand\gnss{\textsc{gnss}\xspace}
\newcommand\gps{\textsc{gps}\xspace}
\newcommand\sbd{\textsc{sbd}\xspace}
\newcommand\spb{\textsc{spb}\xspace}
\newcommand\tdrss{\textsc{tdrss}\xspace}
\newcommand\wifi{\textsc{W}i\textsc{F}i\xspace}
\newcommand\pcb{\textsc{pcb}\xspace}
\newcommand\cpu{\textsc{cpu}\xspace}

\newcommand\cpe{\textsc{cpe}\xspace}
\newcommand\pla{\textsc{pla}\xspace}
\newcommand\petg{\textsc{petg}\xspace}
\newcommand\sd{\textsc{sd}\xspace}
\newcommand\usb{\textsc{usb}\xspace}

\newcommand\utias{{University of Toronto Institute for Aerospace Studies (UTIAS), 4925 Dufferin Street, Toronto, ON, Canada}}

\newcommand\princeton{{Department of Physics, Princeton University, Jadwin Hall, Princeton, NJ 08544, USA}}
\newcommand\durham{{Department of Physics, Durham University, South Road, Durham DH1 3LE, UK}}

\newcommand\uoftastro{{Dunlap Institute for Astronomy and Astrophysics, University of Toronto, 50 St. George Street, Toronto, ON M5S 3H4, Canada}}
\newcommand\uoftphysics{{Department of Physics, University of Toronto, 60 St. George Street, Toronto, ON, Canada}}
\newcommand\jpl{{Jet Propulsion Laboratory, California Institute of Technology, 4800 Oak Grove Drive, Pasadena, CA, USA}}

\newcommand\washu{{Department of Physics, Washington University in St.\ Louis, 1 Brookings Drive, St.\ Louis, MO 63130, USA}}

\newcommand\sydney{{School of Physics, The University of Sydney and ARC Centre of Excellence for Dark Matter Particle Physics, NSW 2006, Australia}}
\newcommand\asiaa{{Academia Sinica Institute of Astronomy \& Astrophysics, No.\ 1, Sec.\ 4, Roosevelt Road, Taipei 10617, Taiwan}}
\newcommand\northeastern{{Department of Physics, Northeastern University, 360 Huntington Ave, Boston, MA USA}}
\newcommand\epfl{{Laboratoire d’Astrophysique, EPFL, Observatoire de Sauverny, 1290 Versoix, Switzerland}}
\newcommand\uoftengsci{{Division of Engineering Science, 40 St. George Street, Room 2110, Toronto ON M5S 2E4, Canada}}
\newcommand\starspec{{StarSpec Technologies Inc., Unit C-5, 1600 Industrial Road, Cambridge, ON N3H 4W5, Canada}}
\newcommand\pltr{{Palantir Technologies, 1875 Lawrence St, Denver, CO 80202, USA}}
\newcommand\caltech{{California Institute of Technology, 1200 E.\ California Blvd., CA 91125, USA}}
\newcommand\csbf{{Columbia Scientific Balloon Facility, 1510 Farm Road 3224,
Palestine, TX 75803, USA}}
\newcommand\wallops{{NASA Wallops Flight Facility, Balloon Program Office, 34200 Fulton St, Wallops Island, VA 23337, USA}}

\newcommand\ndurham{1}
\newcommand\nsydney{2}

\newcommand\nuoftastro{3}

\newcommand\nprinceton{4}
\newcommand\njpl{5}
\newcommand\ncsbf{6}
\newcommand\nstarspec{7}
\newcommand\nepfl{8}
\newcommand\nutias{9}
\newcommand\nwallops{10}
\newcommand\nuoftphysics{11}
\newcommand\nnortheastern{12}

\newcommand\nwashu{13}

\newcommand\nuoftengsci{14}
\newcommand\npltr{15}
\newcommand\ncaltech{16}
\newcommand\nasiaa{17}

\Title{Data downloaded via parachute \texorpdfstring{\\}{}from a NASA super-pressure balloon}

\TitleCitation{Title}

\Author{%
Ellen L.\ Sirks$^{\ndurham,\nsydney,*,\dagger}$\orcidA{}, 
Richard Massey$^{\ndurham,*,\dagger}$\orcidB{},
Ajay S.\ Gill$^{\nuoftastro}$\orcidC{},
Jason Anderson$^{\ndurham}$,
Steven J.\ Benton$^{\nprinceton}$\orcidD{},
Anthony M.\ Brown$^{\ndurham}$\orcidE{},
Paul Clark$^{\ndurham}$\orcidF{},
Joshua English$^{\ndurham}$,
Spencer W.\ Everett$^{\njpl}$\orcidG{},
Aurelien A.\ Fraisse$^{\nprinceton}$,
Hugo Franco$^{\ncsbf}$,
John W.\ Hartley$^{\nstarspec}$,
David Harvey$^{\nepfl}$\orcidH{},
Bradley Holder$^{\nutias}$,
Andrew Hunter$^{\ndurham}$,
Eric M.\ Huff$^{\njpl}$\orcidI{},
Andrew Hynous$^{\nwallops}$,
Mathilde Jauzac$^{\ndurham}$\orcidJ{},
William C.\ Jones$^{\nprinceton}$\orcidK{},
Nikky Joyce$^{\ndurham}$,
Duncan Kennedy$^{\ndurham}$,
David Lagattuta$^{\ndurham}$\orcidL{},
Jason S.-Y.\ Leung$^{\nuoftastro}$\orcidM{},
Lun Li$^{\nprinceton,\nstarspec}$\orcidN{},
Stephen Lishman$^{\ndurham}$,
Thuy Vy T.\ Luu$^{\nprinceton}$,
Jacqueline E.\ McCleary$^{\nnortheastern}$\orcidO{},
Johanna M.\ Nagy$^{\nwashu}$\orcidP{},
C.\ Barth Netterfield$^{\nuoftastro,\nuoftphysics}$,
Emaad Paracha$^{\nuoftphysics}$\orcidQ{},
Robert Purcaru$^{\nuoftengsci}$,
Susan F.\ Redmond$^{\nprinceton}$\orcidR{},
Jason D.\ Rhodes$^{\njpl}$\orcidS{},
Andrew Robertson$^{\njpl}$,
L.\ Javier Romualdez$^{\nstarspec}$,
Sarah Roth$^{\nwallops}$\orcidZ{},
Robert Salter$^{\ncsbf}$,
J\"urgen Schmoll$^{\ndurham}$\orcidT{},
Mohamed M.\ Shaaban$^{\nuoftphysics,\npltr}$\orcidU{},
Roger Smith$^{\ncaltech}$\orcidY{},
Russell Smith$^{\njpl}$\orcidX{},
Sut Ieng Tam$^{\nasiaa}$\orcidV{},
Georgios N.\ Vassilakis$^{\nnortheastern}$\orcidW{}
}

\AuthorNames{Ellen Sirks, Richard Massey et al}

\AuthorCitation{Sirks, E.\ L.; Massey, R.; et al.}

\address{%
$^{\ndurham}$ \quad \durham\\
$^{\nsydney}$ \quad \sydney\\
$^{\nuoftastro}$ \quad \uoftastro{}\\
$^{\nprinceton}$ \quad \princeton\\
$^{\njpl}$ \quad \jpl\\
$^{\ncsbf}$ \quad \csbf\\
$^{\nstarspec}$ \quad \starspec\\
$^{\nepfl}$ \quad \epfl\\
$^{\nutias}$ \quad \utias\\
$^{\nwallops}$ \!\!\quad \wallops\\
$^{\nuoftphysics}$ \!\!\quad \uoftphysics\\
$^{\nnortheastern}$ \!\!\quad \northeastern\\
$^{\nwashu}$ \!\!\quad \washu\\
$^{\nuoftengsci}$ \!\!\quad \uoftengsci\\
$^{\npltr}$ \!\!\quad \pltr\\
$^{\ncaltech}$ \!\!\quad \caltech\\
$^{\nasiaa}$ \!\!\quad \asiaa\\
}

\corres{Correspondence: ellen.sirks@sydney.edu.au, r.j.massey@durham.ac.uk}

\firstnote{These authors contributed equally to this work.} 

\abstract{
In April to May 2023, the \superbit telescope was lifted to the Earth’s stratosphere by a helium-filled super-pressure balloon, to acquire astronomical imaging from above (99.5\% of) the Earth's atmosphere. It was launched from New Zealand then, for 40 days, circumnavigated the globe five times at a latitude 40 to 50 degrees South. Attached to the telescope were four ``\drs'' (Data Recovery System) capsules containing 5\,TB solid state data storage, plus a \gnss receiver, Iridium transmitter, and parachute. Data from the telescope were copied to these, and two were dropped over Argentina. They drifted 61\texorpdfstring{\,}{}km horizontally while they descended 32\texorpdfstring{\,}{}km, but we predicted their descent vectors within 2.4\texorpdfstring{\,}{}km: in this location, the discrepancy appears irreducible below \texorpdfstring{$\sim$}{}2\texorpdfstring{\,}{}km because of high speed, gusty winds and local topography. The capsules then reported their own locations to within a few metres. We recovered the capsules and successfully retrieved all of \superbit's data --- despite the telescope itself being later destroyed on landing.
}

\keyword{Balloon instrumentation; Data handling; Data compression; Models and simulations; Large detector-systems performance} 

\begin{document}


\section{Introduction}

The Superpressure Balloon-borne Imaging Telescope \citep[\superbit;][]{Romualdez_2016,Romualdez_2018} was raised to ${\sim}$33\,km altitude by \nasa's 18.8~million cubic foot super-pressure balloon 728NT\footnote{\url{https://www.csbf.nasa.gov/map/balloon10/flight728NT.htm}}. It was launched on 16~April~2023 from W\={a}naka, New Zealand, was carried East by seasonal polar vortex winds \citep{Karoly_2015}, circumnavigated the Earth ${\sim}$5.5~times, then landed in Argentina on 25~May~2023. During this flight, the telescope acquired deep optical and near-UV imaging of galaxy clusters and other astronomical objects \citep{Shaaban_2022,McCleary_2023,2020Gill}. 
Early in the mission, data were transmitted to ground stations via Starlink and \tdrss satellite communications links. However, the Starlink connection was lost on 1~May~2023, and \tdrss became unstable on 24~May~2023.  The root cause for each failure remains under investigation.  All of the data were stored on board, as well as on redundant sets of solid state drives within drop packages.  Several copies of the full data set were released via parachute prior to termination of the flight, and were safely recovered.  The \superbit payload was entirely destroyed on landing, and although in this case the data were eventually recovered from the remains of the on board drives, the use of parachutes to ensure the safe recovery of the data proved invaluable.

Four Data Recovery System (\drs) capsules were attached to the telescope before launch. As well as a parachute, each included 5\,TB solid state data storage, a Global Navigation Satellite System (\gnss) receiver so it could work out its location, and an Iridium Short-Burst Data (\sbd) transceiver, so it could report its location to the recovery team \citep[and receive commands to, e.g., turn on or off a beeper][]{Clark_2019}. Two \drss were released, and both were successfully recovered, with data intact.

In this article, we describe upgrades (since the first test of a \drs during a \superbit engineering flight in 2019) to their release mechanism, communications with the telescope, and thermal control. We describe the performance of that hardware during operations, and also the performance of software that we use to predict their decent trajectory. Accurate predictions of the descent vector are essential to chose a safe landing zone. Neither the trajectory of the main telescope payload nor the parachute are actively controlled, but the timing of the release can be chosen so that the \drs lands in an area that is unpopulated for safety, but near a road to facilitate recovery.

Our hardware design\footnote{\url{https://github.com/PaulZC/Data_Recovery_System}} and software\footnote{\url{https://github.com/EllenSirks/pyBalloon}} are both open source. Their success inspired \nasa's `{\sc float}ing {\sc dragon}' Balloon Challenge\footnote{\url{https://floatingdragon.nianet.org/}}. Following this programme, a similar capability to download data or small physical samples will be made available by \nasa to all future super-pressure balloon teams, to insure against loss of communications, loss of payload, or destruction of payload on landing. Our experience suggests that teams should take up the offer.

\begin{figure}[t]
\centering
\includegraphics[width=\linewidth]{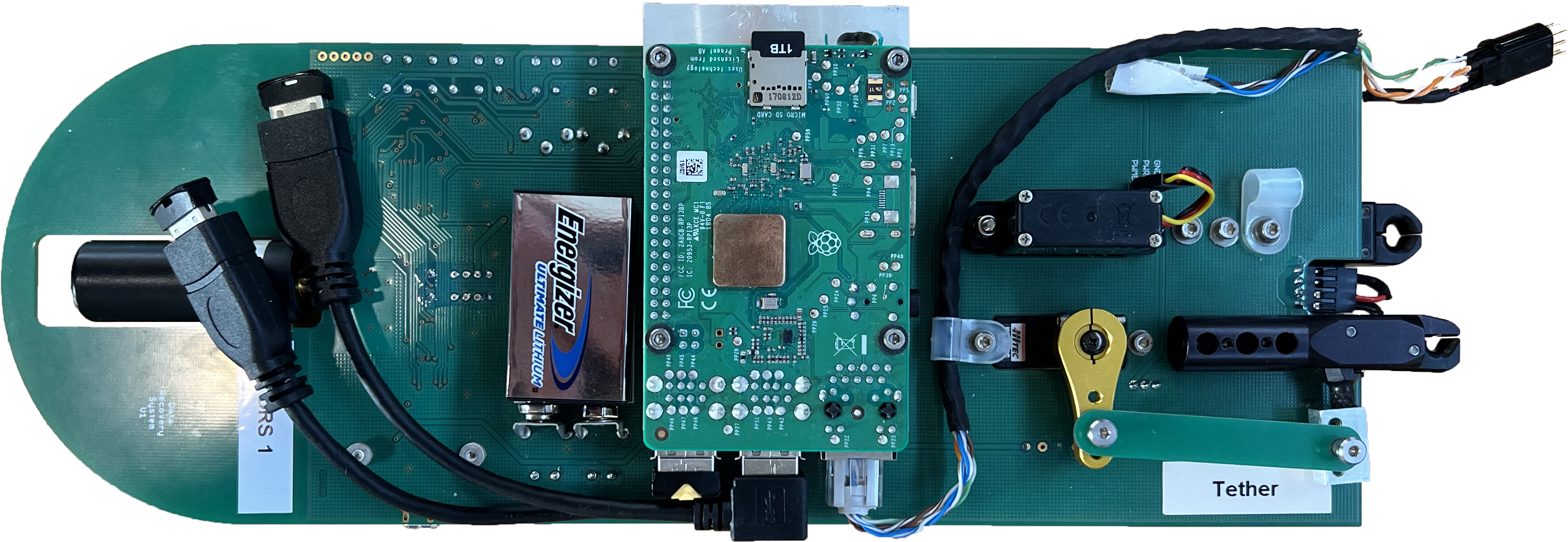}
\caption{The \drs is based around this custom \pcb with a Raspberry Pi in the middle. An Ethernet cable is plugged into the bottom right of the Pi, and attaches to a zero-extraction force connector at the top right corner of the \pcb. Two \sd card readers are plugged directly into the Pi, and two are attached to \usb extension cables to reduce heat production at the sockets. Servo-operated pincer mechanisms can be seen on the right, with the pincer on the front holding on to the main gondola, and the release at the back holding the parachute.}
\label{fig:drs_pcb}
\end{figure}

\begin{figure}[b]
\centering
\includegraphics[width=\linewidth]{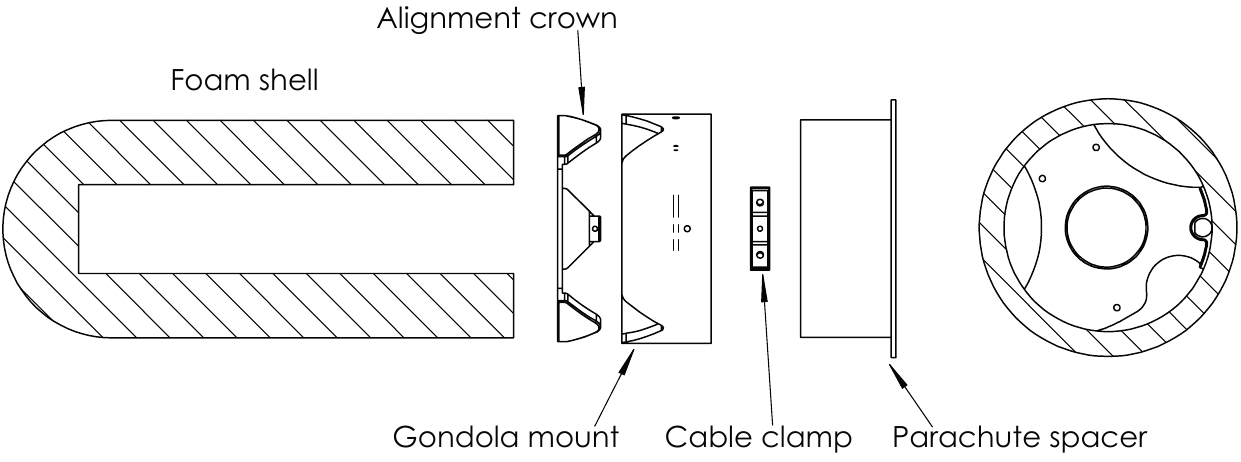}
\caption{The \drs capsules hang underneath the payload (here down is shown to the left). They are prevented from swinging or rotating by a 3D printed `alignment crown' fixed to their top --- which fits inside a mount on the gondola that has the inverted shape.}
\label{fig:crown}
\end{figure}  

\section{Upgraded hardware design}

The \drs capsules are based around a Raspberry Pi 3B attached to a custom 300\,mm $\times$ 125\,mm printed circuit board (\pcb; see figure~\ref{fig:drs_pcb}). They are powered by switchable 24V DC from the main payload while they are attached, then by two internal 9V lithium batteries after release. {These batteries are the only components of the \drs that could be considered potentially hazardous, however, they are compliant with safety test criteria T1--T8 defined in section~38.3 of \cite{UN_manual}}. They use a geared pincer mechanism to hold on to a loop of nylon cable tie underneath the main payload. After coordination with local Air Traffic Control, we issue a sequence of three commands (to prevent errors and fault propagation) that use a servo inside the \drs to open the pincer, allowing the \drs to drop away. 

Each \drs is enclosed by a 3D-printed \cpe hard plastic case and a closed-cell foam shell (see figure~\ref{fig:crown}). One case was printed in \pla, which also worked fine but is harder to work because it is brittle. The foam shell absorbs impacts and adds a degree of waterproofing that proved useful when one \drs landed in snow. It is manufactured by pouring liquid PU expanding foam mix into a pill-shaped mould a few millimetres less than 6'' in diameter, lined with a chicken roasting bag to create a waterproof outer surface on which contact details can be written \citep{Sirks_2020}.  The total mass including parachute is 1.28\,kg. 


\subsection{Alignment and attachment to the main payload}

To prevent the \drs swinging while it hangs under the payload, during the 2019 test flight, it was sheathed inside a section of 6'' plastic drainpipe. This worked flawlessly at altitude, when two \drs capsules were successfully released. However, during subsequent testing in a thermal vacuum chamber, the foam sometimes expanded and became wedged inside the pipe --- to drop only when surrounding air pressure increased. To mitigate the risk of a \drs becoming wedged at altitude, we remove the pipe, and attach a 3D-printed {\sc petg} `crown' to the top of the \drs (see figure~\ref{fig:crown}). This fits into an inverted crown that is fixed to the main gondola. The crowns are kept together and the \drs held secure by pulling the cable tie taught.

A second pincer mechanism holds a 4\,foot parachute, folded inside a spacer above the \drs. This size parachute yields a terminal velocity $\sim$4\,m\,s$^{-1}$. This is fast enough to reduce downrange travel distance (hence uncertainty in predicted landing spot, which allows us to avoid populated areas), while keeping kinetic energy comfortably below a 15\,J safety requirement not to cause serious injury or damage.

\subsection{Hardwired Ethernet to the main payload}

It is possible for the \drs to communicate with the main payload via the Raspberry Pi's built-in \wifi. This worked successfully throughout the 2019 test flight --- however, the launch of that flight was delayed because of radio interference. After much investigation, it turned out that the interference had nothing to do with \wifi, but it took time to rule out that possibility and to then look for the real source. To militate against future risk of delays, and to enable operation on payloads that require radio silence for scientific measurements \citep[e.g.\ SPIDER;][]{Gualtieri_2018}, we use hardwired Ethernet for the 2023 science flight. Inside the \drs, this involves a standard Ethernet cable plugged into the Pi's Ethernet socket. Outside the \drs, two twisted cable pairs were routed to a switch on \superbit, which acted as a {\sc dns} server.

The six wires (four for Ethernet and two for power) join at zero-extraction force connectors\footnote{Smiths Interconnect Hypertac\textsuperscript{\textregistered} connectors, part numbers D02PB906FSTAH and D02PB906MSTH.} that separate easily when the \drs is released. 2-pin versions of these connectors worked well during the test flight, but we found that versions with more pins did not sit well, and occasionally became disconnected during testing. We therefore used grub screws to fix half of the solid connector inside the crown, and half in the gondola mount, so they would be held together until release.

\begin{figure}[t]
\centering
\includegraphics[width=1\linewidth]{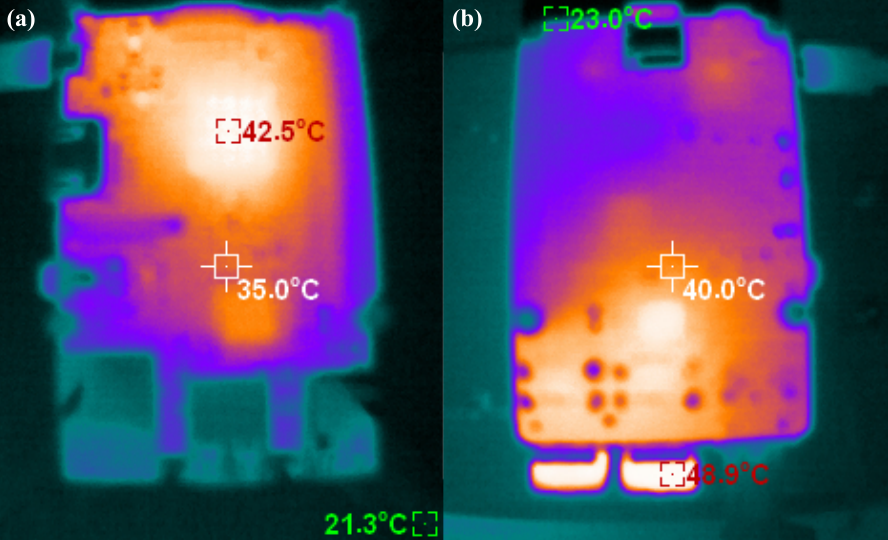}
\caption{Thermal imaging of the top and bottom of a Raspberry Pi, during testing in a room temperature laboratory. Thermal emission is interpreted as temperature (red, white, green numbers) assuming emissivity $e=0.95$. \textbf{(a)} The main \cpu is the hottest component while executing a simple Python script. Nothing is attached to the \usb sockets in this image. \textbf{(b)} The underside of a Pi attached to a \drs, oriented as in figure~\ref{fig:drs_pcb}. The heat sink and \sd card containing the Operating System are now cold at the top, but the \usb sockets become hot during file transfer to \sd cards in small readers mounted in the \usb sockets.}
\label{fig:rpi_temp}
\end{figure}

\subsection{Thermal management} \label{sec:thermal}

During the 2019 engineering flight, each \drs worked for only $\sim$15\,minutes at a time, before overheating and shutting down until it naturally cooled. 
Subsequent testing in a room temperature lab showed two main hot spots (see figure~\ref{fig:rpi_temp}). Assuming thermal emissivity coefficient $e=0.95$, the main \cpu reaches $42.5^{\circ}$\,C while executing a simple Python script (the Raspberry Pi itself was reporting a \cpu temperature of $49.4^{\circ}$\,C), and the \usb sockets reach $48.9^{\circ}$\,C when a 5\,GB file is written to an \sd card in a small \sd card reader. Achieving robust operation during the 2023 science flight required mitigations for both.

To cool the main \cpu, we mill a 53\,gram aluminium heat sink/radiator that sits between the Raspberry Pi and the \pcb. It is held against the \cpu by a spring glued to the \pcb, and contact is ensured with thermal paste. The aluminium block extends through a 56\,mm$\times$14\,mm hole in the plastic case, ending flush with that, and is covered with Mylar tape to increase its infrared emissivity. Heat is radiated to space through a hole in the foam shell that was cut by hand (using an indentation in the mould to align it). The ambient temperature in the stratosphere is $-40^\circ$C \cite{Oliver1987}. For what it is worth, we also stick a thin, off-the-shelf copper radiator to the {\sc ram} chip on the back. 

To reduce heat production at the \usb sockets, we use two 6'' \usb extension cables to relocate two \sd card readers (they end up sitting roughly behind the microcontroller). 

\begin{figure}[t]
\centering
\includegraphics[width=\linewidth]{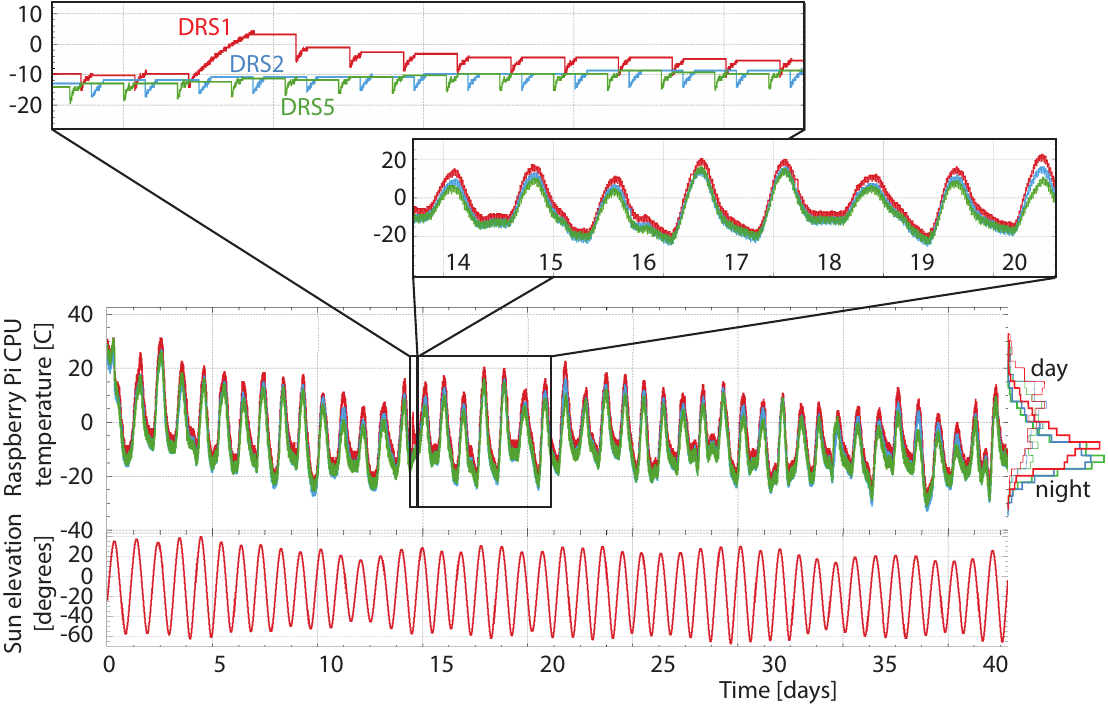}
\caption{Self-reported \cpu temperature of the Raspberry Pi on each \drs during the \superbit mission, for \drsi (red), \drsii (blue), and \drsv (green). Temperature principally varies with the 45 diurnal cycles in 40\,days (\superbit crossed the international date line 5 times), the length and extent of which varied with \superbit's speed and latitude. The histogram on the right splits times by positive or negative Solar elevation. Insets show temperature variations with increased temporal resolution --- including, for illustration, the one occasion when \drsi was left powered on for 30\,minutes. Temperature measurements were only available when a \drs was powered on, and horizontal lines merely show the last reported temperature.
}
\label{fig:rpi_temp_inflight}
\end{figure}

We also operate each \drs on a 5\,minute on--20\,minute off rota: powering up each in turn, copying data, then powering down. During a typical cycle during the night (day), the \cpu temperature increases by 3.0 $\pm$ 0.3$^\circ$C (3.7 $\pm$ 1.0$^\circ$C) in 5\,minutes (see figure~\ref{fig:rpi_temp_inflight}). On one occasion during night 13, \drsi was powered for two cycles, and its temperature rose 14.7$^\circ$C (from $-11.6^\circ$C to 3.1$^\circ$C) in 30\,minutes. It continued operating throughout this time, and cooled half-way back to the same temperature as the others in approximately 2\,hours. We do not know why \drsi runs at $-8.2\pm8.6^\circ$C ($4.7\pm11.2^\circ$C) during the night (day), consistently a few degrees warmer than \drsii at $-11.4\pm8.1^\circ$C ($1.0\pm10.7^\circ$C) and \drsv at $-11.9\pm7.4^\circ$C ($-1.2\pm10.6^\circ$C).


\begin{figure}[t]
\centering
\includegraphics[width=\linewidth]{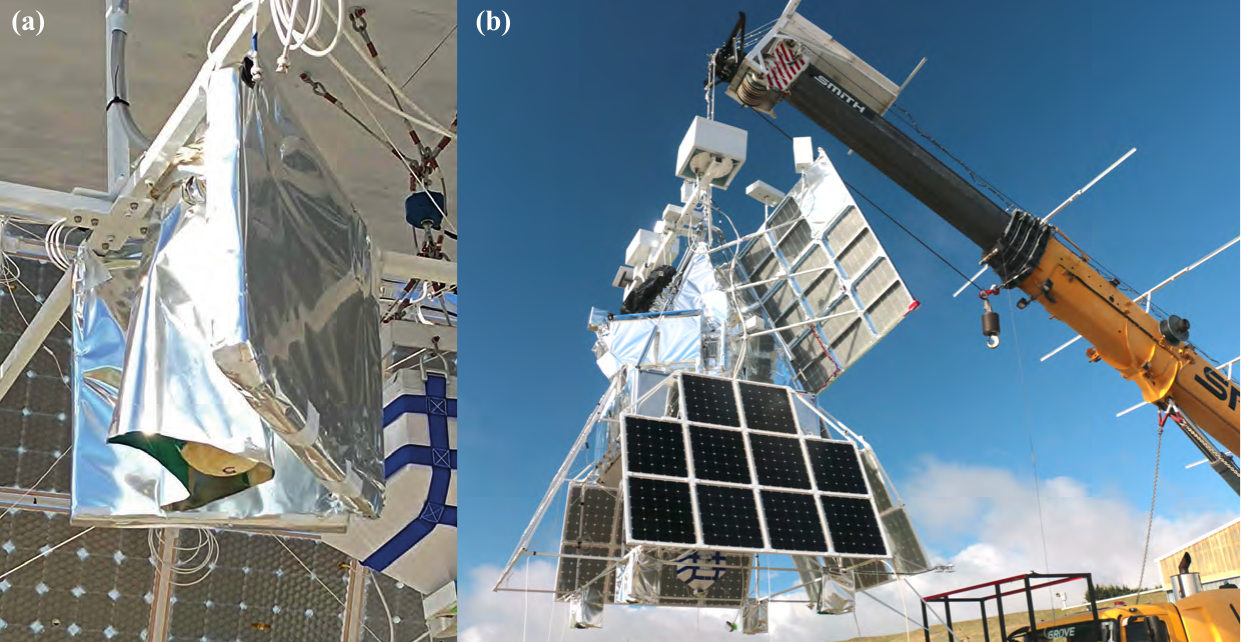}
\caption{\textbf{(a)} Close-up of a \drs in flight configuration. The closed-cell foam shell surrounding the \drs can be seen poking out below a skirt of aluminised Mylar. It is further protected on two sides by sheets of foam insulation also covered by aluminised Mylar. Cables are routed upwards, on the mounting frame. \textbf{(b)} \superbit suspended on the launch crane. Four \drs capsules can be seen at the bottom, each attached to a corner of the frame holding the solar panels. The blue and white object hanging between them is a ballast hopper. Throughout its mission, the telescope keeps its back (on the right in these photos) oriented towards the Sun.}
\label{fig:drs_in_situ}
\end{figure}

Before launch, four \drs capsules were mounted underneath the rest of the payload (see figure~\ref{fig:drs_in_situ}). They are oriented with their radiators towards the front of the telescope (which is never allowed to point towards the Sun) and enclosed by a loose skirt of Mylar, the inside of which is aluminised except for a $\sim$45$^\circ$ pleat near the radiator. On two sides, an additional sunshield of foam wrapped in aluminised Mylar also protects the \drss from incidental knocks. \drsi was at the back left, \drsii at the back right, \drsiii at the front right, \drsv at the front left (\drsiv was our flight spare). Power and Ethernet cables are routed from each \drs up to \superbit.

\subsection{Data storage}

Each \drs is fitted with $5\times1$\,TB MicroSDXC cards: one in the internal slot that includes the Operating System, and four in \usb card readers. We set these up as independent drives (rather than {\sc raid}, to further reduce thermal load), and copy data to one of the \sd cards at a time, beginning with those mounted in a \usb socket. An \texttt{rsync} script was not sufficient, because the \sd cards were independent. Instead, whenever the Pi boots, a custom script compares files on the payload with those on the \drs (matching name and size, because some files such as instrument timestreams continually grow), and copies new data to an \sd card with sufficient space. {\superbit}'s rate of data acquisition is sufficiently slow that we could have duplicated all the data to second \sd cards on every \drs\ --- but with so many backups already, we chose not to.

We tested MicroSDXC cards from several manufacturers before the flight. Resilience is more important than speed in this context, and we found considerable variation in cards' ability to withstand a harsh environment. SanDisk Ultra (A1 UHS-I Class 10 U1) and SanDisk Extreme (A2 UHS-I Class 10 U3 V30) cards worked reliably throughout our tests and the flight. Two different Integral (A2 UHS-I Class 10 U3 V30) cards showed an intermittent fault during testing: when used to house the Operating System, the Pi would occasionally require power cycling to boot. After a few early problems, this was not reproducible, so we flew two \drss with SanDisk Ultra \sd cards and two with Integral \sd cards. All our Lexar (LMSPLAY001T-BNNAG) cards all failed during testing, so these were discarded.

%

\subsection{One failure on launch}

One \drs onboard \superbit stopped responding immediately after launch. It is now impossible to determine why, but we suspect that the zero-extraction force connection became misaligned by the launch shock. It was one of the capsules using an Integral \sd card, which may have also caused problems, but we power cycled it many times without response. All three other \drs capsules worked reliably for the entire flight or until they were released. 

\section{Descent and recovery}

We initially intended to release one \drs capsule around mission days 40, 60, 80, and 100, whenever \superbit passed over land, to mitigate the risk of the payload and all its on-board data being lost at sea. \superbit did indeed pass over Chile and Argentina on mission day 40 (25~May~2023). 
However, in the early hours of the morning it became apparent that the main payload would need to descend later that day, because of reduced communications bandwidth, and a weather forecast indicating a Southerly trajectory away from further land crossings. After finishing the night's observing, and coordinating with local Air Traffic Control, we released (not one but) two \drs capsules over Santa Cruz province, Argentina. This strategy ensured that independent, redundant sets of data would be available both onboard and separated from the main payload. 

\subsection{Release time and location} \label{sec:release_time}

We attempted to release two \drs capsules nearly simultaneously --- so they would land close to each other, to simplify recovery. Low-latency communications with the payload had been lost, so we were forced to send release commands via Iridium \sbd. We transmitted the two commands on independent channels. However, commands are queued and arrive after an unpredictable delay.

Contemporaneous logs recorded on \superbit 
show that release commands were received at 12:30:58 and 12:31:25~UT. Taking into account an intentional 30\,seconds delay for the Pi to shut down gracefully, we infer that \drsii and \drsi were released at 12:31:28 and 12:31:55~UT respectively.

The super-pressure balloon (\spb) host logged \gnss locations approximately once per minute. We interpolate between these to infer the location of \superbit at the moments when \drs capsules were released (see table~\ref{tab:release_times}). During the 27\,seconds between release events, \superbit travelled 1.79\,km on heading 101$^\circ$. 


\begin{table}[H] 
\caption{Contemporaneous logs from on board \superbit, around the time when two \drs capsules were released. Times are UT on 25~May~2023. \spb logs \#1--4 record accurate GPS locations at known times, which we interpolate to the moments of release for \drsii and \drsi.\label{tab:release_times}}
\newcolumntype{M}{>{\centering\arraybackslash}X}
\begin{tabularx}{\textwidth}{lMMMM}
\toprule
\textbf{Event} & \textbf{Time (UT)} & \textbf{Latitude [deg]} & \textbf{Longitude [deg]} & \textbf{Altitude [m]}\\
\midrule

\spb log~\#1     & 12:29:33 & $-47.685$ & $-70.518$ & 32,585\\ 
\spb log~\#2     & 12:30:37 & $-47.692$ & $-70.475$ & 32,488\\ 
Release of \drsii& 12:31:28 & $-47.696$ & $-70.444$ & 32,439\\ 
\spb log~\#3     & 12:31:40 & $-47.697$ & $-70.437$ & 32,428\\ 
Release of \drsi & 12:31:55 & $-47.699$ & $-70.421$ & 32,323\\ 
\spb log~\#4     & 12:32:44 & $-47.707$ & $-70.367$ & 31,981\\ 

\bottomrule
\end{tabularx}
\end{table}



\subsection{Predicted descent trajectories}

We predict the \drs descent trajectories and landing sites with {\tt PyBalloon} software \citep{Sirks_2020} that integrates a path followed by a test particle through winds described by Global Forecast System ({\sc gfs}) weather models.
When we did this live, we predicted descent vectors $\sim$61.3\,km on heading $107^\circ$, and a flight time of $\sim$32\,minutes. 
However, the fast ground speed and high-latency communications meant that our uncertainty in the landing site was dominated by the unpredictable timing of release (2 minutes' uncertainty extended our error ellipses by 8\,km). We therefore aimed for a large region of open, uninhabited~land west of main road 12. These specific circumstances are unusual and unlikely to be repeated.

\begin{figure}[t]
\centering
\includegraphics[width=\linewidth]{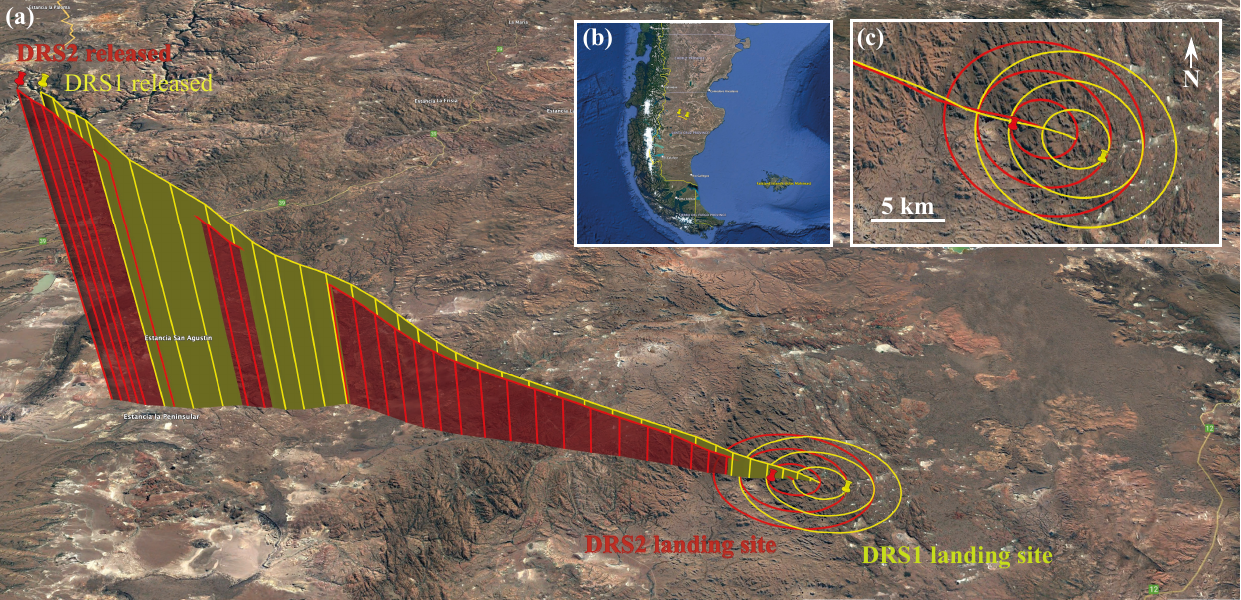}
\caption{Descent trajectories of \drsi (yellow) and \drsii (red) capsules, which were released over southern Argentina on 25~May~2023. Pins mark the known release and landing positions; everything else is modelled using the pyBalloon software \citep{Sirks_2020}. \textbf{(a)} View from altitude, looking North. 
Predicted descent trajectories, spanning 62\,km horizontally and 32\,km vertically are shown, with vertical lines every 15\,seconds for the first minute, then every 1\,minute. \textbf{(b)} Location of trajectory within Argentina. \textbf{(c)} Zoomed-in map view, comparing the true landing sites to the predicted $1\sigma$, $2\sigma$ and $3\sigma$ uncertainty ellipses. The background image was created with Google Earth, using data from SIO, NOAA, U.S. Navy, NGA, GEBCO and imagery from Landsat/Copernicus.}
\label{fig:drs_predictions}
\end{figure}

Knowing with hindsight the precise times of release, we predict landing sites for~\drsi at $-47.8594^\circ$, $-69.6307^\circ$ (altitude 878\,m), and \drsii at $-47.8557^\circ$, $-69.6559^\circ$ (altitude 920\,m), with 68\% confidence limits of $\sigma_\parallel=2.1$\,km in the main direction of motion, and $\sigma_\perp=1.9$\,km perpendicular to that.
With even more hindsight, we are able to interpolate between {\sc gfs} model wind speeds issued before {\it and after} the \drs release. 
These lengthen the model descent vectors by $\sim$1.1\,km, to predicted landing sites for~\drsi at $-47.8655^\circ$, $-69.6153$ (altitude 863\,m) and \drsii at $-47.8600^\circ$, $-69.6444^\circ$ (altitude 850\,m), with 68\% confidence limits $\sigma_\parallel=2.3$\,km and $\sigma_\perp=1.9$\,km (see figure~\ref{fig:drs_predictions}).
Even in the absence of any additional information, the open terrain would have made it possible (albeit time consuming) to search for bright orange parachutes in an area this size. 

\subsection{Reported landing sites}

Unfortunately, neither \drs reported or recorded its location during descent, as they are supposed to (and as they did during the 2019 test flight). It is therefore impossible to reconstruct their true trajectories. \drsi spontaneously started reporting back at 14:13:46~UT, after having reached the ground, and \drsii started reporting back at 14:31:15~UT (intermittently at first, and intermittently again overnight). Both \drs capsules reported line of sight to many \gnss satellites. However, their initial reports included measurements of low battery voltages, and we later found that \drsii was lying in a patch of snow. We infer that, although the temperature of the Raspberry Pis was successfully managed in space for 40 days (see section~\ref{sec:thermal}), the adjacent batteries, which are used only after release, got very cold. We have previously used them in the stratosphere, but not for such extended periods.

Self-reported \gnss coordinates after landing placed \drsi at $-47.88167^\circ$, $-69.59300^\circ$ (altitude 786\,m) and \drsii at $-47.86368^\circ$, $-69.67452^\circ$ (altitude 806\,m), after averaging over several readings to provide $\sim$2\,m precision. These are 66.5\,km and 58.3\,km horizontally from their release points, 3.8\,km and 1.6\,km from the pre-release predicted locations, or 2.4\,km and 2.3\,km from our best-estimate calculations that include weather data available afterwards. The true trajectories appear to have deviated approximately 950\,m sideways from the predictions, with one landing 2.2\,km earlier than expected, and the other travelling 2.1\,km farther: both within predicted precision.

Our model for both the descent vector and its uncertainty thus appear accurate. The absolute uncertainty on our predictions is larger than previous drops because the high wind speed meant that capsules travelled farther downrange. Indeed, we have to extrapolate uncertainties from most of the test-drops used to calibrate our model \citep{Sirks_2020}. But even in this regime, compared to our predicted landing sites, three of the four coordinate offsets are inside the predicted 68\% confidence interval, with one just outside. Compared to our best-guess landing sites, all four offsets are {\it just} inside. 
Furthermore, the two near-simultaneous trajectories neatly illustrate the irreducible component of uncertainty caused by small-scale spatial and temporal variations in wind velocity (gusts) --- as well as local topography, when the predicted trajectory at low altitude is only $21^\circ$ below the horizontal. The two capsules started only 1.8\,km (and 27\,seconds) apart, but ended up separated by 6.4\,km.


\begin{figure}[t]
\centering
\includegraphics[width=\linewidth]{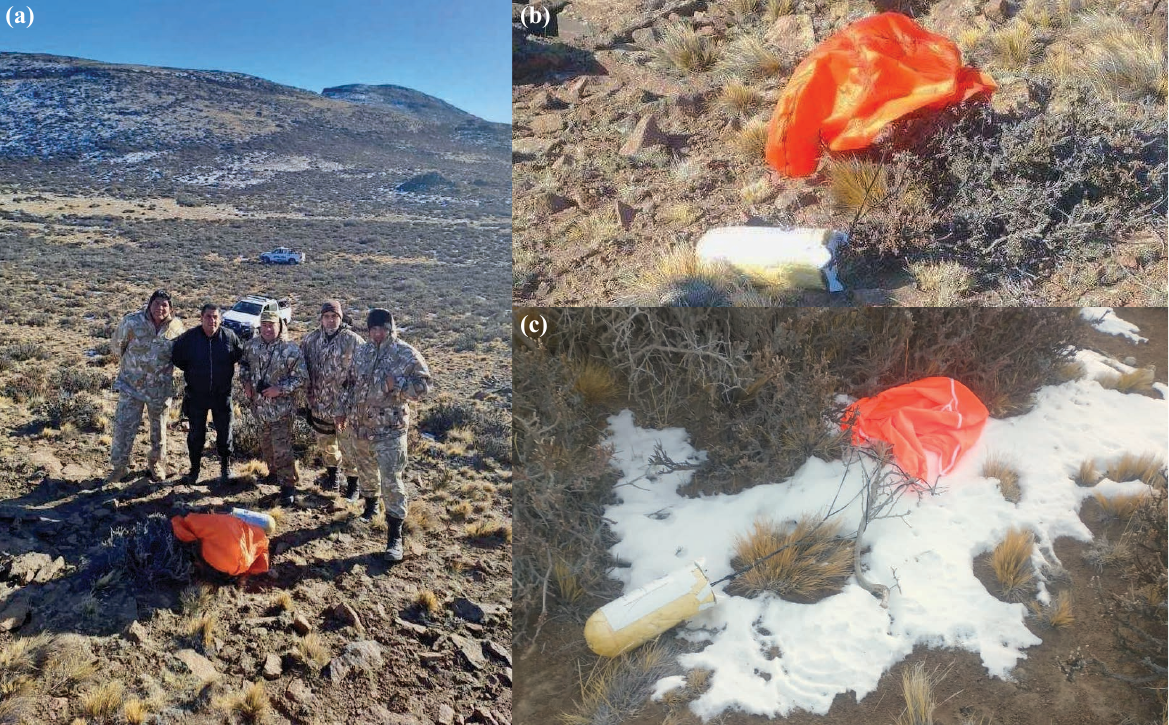}
\caption{Landing sites of the \drs capsules in Argentina. \textbf{(a)} General terrain and the Search and Rescue team from the Governor's Office of Santa Cruz Province. The bright orange parachutes of \drsi \textbf{(b)} and \drsii \textbf{(c)} were visible from distance. The white foam shell and release crown are finally also visible; the foam helped to insulate and waterproof \drsii when it landed on snow.
}
\label{fig:drs_recovery}
\end{figure}

\subsection{Successful recovery}

Both \drs capsules were found at their reported locations in rolling hills (see figure~\ref{fig:drs_recovery}), by the Search and Rescue team for the Governor's Office of Santa Cruz Province. The bright orange parachutes were visible from a distance and the beeping was audible. \drsi was found exactly at its reported location, a few hundred metres from a track, and $\sim$24\,km from main road 12. \drsii was found a few metres from its reported location, with evidence of cougar prints in the snow. We surmise that foam and parachute nylon are intriguing but not tasty.

We recovered identical copies of all the data from both released \drs capsules --- and also later from the unreleased \drs and the main data store on \superbit (which also have slightly more telemetry data). However, \superbit had been completely destroyed on landing, when its parachute failed to detach (perhaps due to similar thermal issues as the \drs capsules: analysis is ongoing) and it was dragged for 3\,km through similar terrain, leaving a trail of debris. It is therefore remarkable luck that \superbit's solid state hard drive was later discovered intact. We did not \textit{need} it, because data had already been retrieved from the released \drs capsules, but having the original copy enabled us to verify that no data on the \sd cards were corrupted.

\section{Results}

The \drs system comprises both hardware and software. Hardware worked reliably throughout the mission, in three of the four capsules. They were power cycled every 15\,minutes for 40\,days, and booted successfully every time. Data were gradually copied to every \drs as they were acquired, and later (after a mean delay of 20\,days at altitude plus two days exposed in Argentina) recovered with no bit errors in 219\,GB of files. Curiously, one of the 5920 files accumulated 10 bit errors when being copied from the telescope's camera data store to the central data store --- but none were further corrupted, on either the central data store or the \drs capsules. 

The internal batteries in the \drs capsules got very cold at altitude, and provided insufficient voltage to power the Iridium transceiver until they warmed up after landing. This meant the true descent trajectories were not recorded --- which would have been interesting to compare to our model trajectories. However, the batteries did warm up, and worked successfully even from a exposed, high altitude location on Earth, while covered in snow. The capsules reported their location and were found.

One of the four \drs capsules failed to respond after launch. Following the destruction of the telescope to which it remained attached, it is impossible to determine the cause. We speculate that a cable or connector may have been dislodged by the launch shock, but we also note that the \drs in question used a brand of \sd card that had displayed erratic behaviour during testing. Either way, only 75\% hardware reliability calls for additional testing before future use, to reduce or improve single points of failure.

The software we developed to predict the trajectory of a \drs descending through the Earth's atmosphere worked very well, with accurate predictions and estimates of uncertainty. This is essential to drop the \drs to a safe but accessible landing site. It would also work as backup information to find a \drs that failed to respond after release (although the reliability of our hardware other than during launch meant that this was not needed).

Analysis of the astronomical data collected by \superbit and the \drs capsules will be presented in forthcoming papers.

\section{Discussion}

The first use of the \drs capsules during a live science mission was a huge success. For a relatively small cost, we insured the scientific returns of \superbit against a loss event \textit{that came true}: high bandwidth communication links failed, then the telescope was destroyed upon landing. We recommend that future balloon missions consider using this or similar systems. Our hardware design and software are open source and freely available. Further development is continuing at \nasa, with plans to offer a facility-class \drs system.

Possible improvements to \drs hardware include (a better way to secure) a low-friction connector so it does not become disconnected during launch. The jerk when a \drs falls away is sufficient that the right amount of rigidity might be provided by a standard Ethernet socket and plug with the retaining lug removed. Power Over Ethernet could remove the need for any separate power cables. A related aspect of this is that securing a \drs to the gondola mount currently requires fiddly routing of cables and cable ties inside a cramped case and pulling the ties taut (see figure~\ref{fig:crown}). More convenient assembly/disassembly would make it easier to perform many more test drops in situ than we achieved. 

Thermal management has greatly improved since our test flight in 2019 \citep{Sirks_2020}, with the Raspberry Pi operating reliably on demand throughout the mission. The internal batteries got cold, and a (throttled) thermal link from the Pi would help them stay ready to work immediately after release. Although our batteries eventually warmed up and worked fine after 40\,days cold, we have not tested whether they would also recover as quickly after 100\,days cold. It would also have been interesting if the \drs stored \gnss data (at high frequency) during descent, even though it was too cold to power the Iridium transmitter, so the trajectory could be compared to our prediction. This is not possible in the current design concept, which intentionally has only a low-power microcontroller chip and no flash memory available after release --- to maximise battery life in case searching for the \drs takes a long time.

%

\vspace{6pt} 



\authorcontributions{
Conceptualization, R.M., L.L., J.R., R.S.\ and R.S.;
methodology, P.C., A.B., A.G., J.E.\ and R.P.; 
software E.S.\ and P.C.; 
investigation (\drs hardware), J.A., A.H., N.J., D.K., S.L.,
investigation (\superbit interface), S.B., S.E., A.F., J.H., D.H., B.H., E.H., M.J., W.J., D.L., J.L., T.V.L., J.M., J.N., C.B.N.,  E.P., S.R., L.J.R., J.S., M.S, S.T.\ and G.V.;
investigation (\spb interface), H.F., A.H., S.R.\ and R.G.S.;
writing, E.S.\ and R.M.;
funding acquisition, R.M.
All authors have read and agreed to the published version of the manuscript.
}

\funding{This research was funded by the Royal Society (grant RGF/EA/180026), the UK Science and Technology Facilities Council (grant ST/V005766/1), and an Impact Acceleration Award to Durham University.  Launch and operational support for the 2023 \superbit flight from W\=anaka, New Zealand, was provided by \nasa. Funding for the development of \superbit was provided by \nasa (grant {\sc apra} NNX16AF65G), the Canadian Institute for Advanced Research ({\sc cifar}), the Canadian Natural Science and Engineering Research Council ({\sc nserc}), and Durham University's astronomy survey fund. This work was done in part at {\sc jpl}, run under a contract for \nasa by Caltech. 
}

\institutionalreview{Not applicable.}

\informedconsent{Not applicable.}

\dataavailability{Descent trajectories are available from \url{https://github.com/EllenSirks/pyBalloon}.
Full instructions to build a \drs are available from \url{https://github.com/PaulZC/Data_Recovery_System}.} 

\acknowledgments{
We thank the Ministerio de Seguridad de la Provincia Santa Cruz, Argentina, for deploying their Search and Rescue team to recover the \drs capsules. The search was organised by Ministro de Seguridad Sr.\ Luca Pratti and Subsecretario de Protecci\'on Civil, Sr.\ Diego Farias. Thank you to Juan Perez, German Alejandro Reynoso, Marcelo Alejandro Pemini and Alan Gabriel Arbe for going into the field --- and sorry for keeping you awake when we accidentally left one \drs beeping overnight.}

\conflictsofinterest{The authors declare no conflict of interest. The funders had no role in the design of the study; in the collection, analyses, or interpretation of data; in the writing of the manuscript; or in the decision to publish the results.} 


\abbreviations{Abbreviations}{
The following abbreviations are used in this manuscript:\\

\noindent 
\begin{tabular}{@{}ll}
\cpe & Co-polyester, a branched version of \petg\\
\drs & Data Recovery System\\
{\sc gfs} & Global Forecast System (weather model)\\
\gnss & Global Navigation Satellite System, of which \gps is a subset\\
\gps & Global Positioning System\\
\pcb & Printed Circuit Board\\
\petg & Glycol-modified polyethylene terephthalate hard plastic\\
\pla & Polylactic acid hard plastic\\
\sbd & Short Burst Data (Iridium satellite message)\\
\spb & Super-Pressure (closed, helium-filled) Balloon\\
\superbit & Super-pressure Balloon-borne Imaging Telescope\\
{\sc tdrss} & Tracking and Data Relay Satellite System
\end{tabular}
}

\newpage




\begin{adjustwidth}{-\extralength}{0cm}

\reftitle{References}


\bibliography{bibliography}

\begin{thebibliography}{999}

\bibitem[{Romualdez} \em{et~al.}(2016){Romualdez}, {Benton}, {Clark},
  {Damaren}, {Eifler}, {Fraisse}, {Galloway}, {Hartley}, {Jones}, {Li},
  {Lipton}, {Luu}, {Massey}, {Barth Netterfield}, {Padilla}, {Rhodes}, and
  {Schmoll}]{Romualdez_2016}
{Romualdez}, L.J.; {Benton}, S.J.; {Clark}, P.; {Damaren}, C.J.; {Eifler}, T.;
  {Fraisse}, A.A.; {Galloway}, M.N.; {Hartley}, J.W.; {Jones}, W.C.; {Li}, L.;
  et~al.
\newblock {The design and development of a high-resolution visible-to-near-UV
  telescope for balloon-borne astronomy: SuperBIT}.
\newblock {\em arXiv e-prints} {\bf 2016}, p. arXiv:1608.02502,
  \href{http://xxx.lanl.gov/abs/1608.02502}{{\normalfont
  [arXiv:astro-ph/1608.02502]}}.

\bibitem[{Romualdez} \em{et~al.}(2018){Romualdez}, {Benton}, {Brown}, {Clark},
  {Damaren}, {Eifler}, {Fraisse}, {Galloway}, {Hartley}, {Jauzac}, {Jones},
  {Li}, {Luu}, {Massey}, {Mccleary}, {Netterfield}, {Redmond}, {Rhodes},
  {Schmoll}, and {Tam}]{Romualdez_2018}
{Romualdez}.; {Benton}, S.J.; {Brown}, A.M.; {Clark}, P.; {Damaren}, C.J.;
  {Eifler}, T.; {Fraisse}, A.A.; {Galloway}, M.N.; {Hartley}, J.W.; {Jauzac},
  M.;  et~al.
\newblock {Overview, design, and flight results from SuperBIT: a
  high-resolution, wide-field, visible-to-near-UV balloon-borne astronomical
  telescope}.
\newblock {\em SPIE} {\bf 2018}, {\em 10702},~arXiv:1807.02887,
  \href{http://xxx.lanl.gov/abs/1807.02887}{{\normalfont
  [arXiv:astro-ph.IM/1807.02887]}}.
\newblock {\url{https://doi.org/10.1117/12.2307754}}.

\bibitem[{Karoly, D}(2015)]{Karoly_2015}
{Karoly, D}., Ed.
\newblock {\em Meteorology of the southern hemisphere}; American Meteorological
  Society: Springer,  2015.

\bibitem[{Shaaban} \em{et~al.}(2022){Shaaban}, {Gill}, {McCleary}, {Massey},
  {Benton}, {Brown}, {Damaren}, {Eifler}, {Fraisse}, {Everett}, {Galloway},
  {Henderson}, {Holder}, {Huff}, {Jauzac}, {Jones}, {Lagattuta}, {Leung}, {Li},
  {T. Luu}, {Nagy}, {Netterfield}, {Redmond}, {Rhodes}, {Robertson}, {Schmoll},
  {Sirks}, and {Sivanandam}]{Shaaban_2022}
{Shaaban}, M.M.; {Gill}, A.S.; {McCleary}, J.; {Massey}, R.J.; {Benton}, S.J.;
  {Brown}, A.M.; {Damaren}, C.J.; {Eifler}, T.; {Fraisse}, A.A.; {Everett}, S.;
   et~al.
\newblock {Weak Lensing in the Blue: A Counter-intuitive Strategy for
  Stratospheric Observations}.
\newblock {\em \aj} {\bf 2022}, {\em 164},~245,
  \href{http://xxx.lanl.gov/abs/2210.09182}{{\normalfont
  [arXiv:astro-ph.IM/2210.09182]}}.
\newblock {\url{https://doi.org/10.3847/1538-3881/ac9b1c}}.

\bibitem[{McCleary} \em{et~al.}(2023){McCleary}, {Everett}, {Shaaban}, {Gill},
  {Vassilakis}, {Huff}, {Massey}, {Benton}, {Brown}, {Clark}, {Holder},
  {Fraisse}, {Jauzac}, {Jones}, {Lagattuta}, {Leung}, {Li}, {T. Luu}, {Nagy},
  {Netterfield}, {Paracha}, {Redmond}, {Rhodes}, {Schmoll}, {Sirks}, and
  {Tam}]{McCleary_2023}
{McCleary}, J.E.; {Everett}, S.W.; {Shaaban}, M.M.; {Gill}, A.S.; {Vassilakis},
  G.N.; {Huff}, E.M.; {Massey}, R.J.; {Benton}, S.J.; {Brown}, A.M.; {Clark},
  P.;  et~al.
\newblock {Lensing in the Blue. II. Estimating the Sensitivity of Stratospheric
  Balloons to Weak Gravitational Lensing}.
\newblock {\em \aj} {\bf 2023}, {\em 166},~134,
  \href{http://xxx.lanl.gov/abs/2307.03295}{{\normalfont
  [arXiv:astro-ph.IM/2307.03295]}}.
\newblock {\url{https://doi.org/10.3847/1538-3881/ace7ca}}.

\bibitem[{Gill} \em{et~al.}(2020){Gill}, {Benton}, {Brown}, {Clark}, {Damaren},
  {Eifler}, {Fraisse}, {Galloway}, {Hartley}, {Holder}, {Huff}, {Jauzac},
  {Jones}, {Lagattuta}, {Leung}, {Li}, {Luu}, {Massey}, {McCleary}, {Mullaney},
  {Nagy}, {Netterfield}, {Redmond}, {Rhodes}, {Romualdez}, {Schmoll},
  {Shaaban}, {Sirks}, {Sivanandam}, and {Tam}]{2020Gill}
{Gill}, A.; {Benton}, S.J.; {Brown}, A.M.; {Clark}, P.; {Damaren}, C.J.;
  {Eifler}, T.; {Fraisse}, A.A.; {Galloway}, M.N.; {Hartley}, J.W.; {Holder},
  B.;  et~al.
\newblock {Optical Night Sky Brightness Measurements from the Stratosphere}.
\newblock {\em \aj} {\bf 2020}, {\em 160},~266,
  \href{http://xxx.lanl.gov/abs/2010.05145}{{\normalfont
  [arXiv:astro-ph.IM/2010.05145]}}.
\newblock {\url{https://doi.org/10.3847/1538-3881/abbffb}}.

\bibitem[Clark \em{et~al.}(2019)Clark, Funk, Funk, Funk, Meadows, Brown, Li,
  Massey, and Netterfield]{Clark_2019}
Clark, P.; Funk, M.; Funk, B.; Funk, T.; Meadows, R.; Brown, A.; Li, L.;
  Massey, R.; Netterfield, C.
\newblock An open source toolkit for the tracking, termination and recovery of
  high altitude balloon flights and payloads.
\newblock {\em Journal of Instrumentation} {\bf 2019}, {\em 14},~P04003.
\newblock {\url{https://doi.org/10.1088/1748-0221/14/04/P04003}}.

\bibitem[{UN. Committee of Experts on the Transport of Dangerous
  Goods}(2019)]{UN_manual}
{UN. Committee of Experts on the Transport of Dangerous Goods}.
\newblock Recommendations on the Transport of Dangerous Goods. Manual of tests
  and criteria.
\newblock Technical report, United Nations, Geneva,  2019.
\newblock \url{https://digitallibrary.un.org/record/3846833}.

\bibitem[{Sirks} \em{et~al.}(2020){Sirks}, {Clark}, {Massey}, {Benton},
  {Brown}, {Damaren}, {Eifler}, {Fraisse}, {Frenk}, {Funk}, {Galloway}, {Gill},
  {Hartley}, {Holder}, {Huff}, {Jauzac}, {Jones}, {Lagattuta}, {Leung}, {Li},
  {Luu}, {McCleary}, {Nagy}, {Netterfield}, {Redmond}, {Rhodes}, {Romualdez},
  {Schmoll}, {Shaaban}, and {Tam}]{Sirks_2020}
{Sirks}, E.L.; {Clark}, P.; {Massey}, R.J.; {Benton}, S.J.; {Brown}, A.M.;
  {Damaren}, C.J.; {Eifler}, T.; {Fraisse}, A.A.; {Frenk}, C.; {Funk}, M.;
  et~al.
\newblock Download by parachute: retrieval of assets from high altitude
  balloons.
\newblock {\em Journal of Instrumentation} {\bf 2020}, {\em 15},~P05014,
  \href{http://xxx.lanl.gov/abs/2004.10764}{{\normalfont
  [arXiv:astro-ph.IM/2004.10764]}}.
\newblock {\url{https://doi.org/10.1088/1748-0221/15/05/P05014}}.

\bibitem[{Gualtieri} \em{et~al.}(2018){Gualtieri}, {Filippini}, {Ade}, {Amiri},
  {Benton}, {Bergman}, {Bihary}, {Bock}, {Bond}, {Bryan}, {Chiang}, {Contaldi},
  {Dor{\'e}}, {Duivenvoorden}, {Eriksen}, {Farhang}, {Fissel}, {Fraisse},
  {Freese}, {Galloway}, {Gambrel}, {Gandilo}, {Ganga}, {Gramillano},
  {Gudmundsson}, {Halpern}, {Hartley}, {Hasselfield}, {Hilton}, {Holmes},
  {Hristov}, {Huang}, {Irwin}, {Jones}, {Kuo}, {Kermish}, {Li}, {Mason},
  {Megerian}, {Moncelsi}, {Morford}, {Nagy}, {Netterfield}, {Nolta},
  {Osherson}, {Padilla}, {Racine}, {Rahlin}, {Reintsema}, {Ruhl}, {Runyan},
  {Ruud}, {Shariff}, {Soler}, {Song}, {Trangsrud}, {Tucker}, {Tucker},
  {Turner}, {List}, {Weber}, {Wehus}, {Wiebe}, and {Young}]{Gualtieri_2018}
{Gualtieri}, R.; {Filippini}, J.P.; {Ade}, P.A.R.; {Amiri}, M.; {Benton}, S.J.;
  {Bergman}, A.S.; {Bihary}, R.; {Bock}, J.J.; {Bond}, J.R.; {Bryan}, S.A.;
  et~al.
\newblock {SPIDER: CMB Polarimetry from the Edge of Space}.
\newblock {\em Journal of Low Temperature Physics} {\bf 2018}, {\em
  193},~1112--1121,  \href{http://xxx.lanl.gov/abs/1711.10596}{{\normalfont
  [arXiv:astro-ph.CO/1711.10596]}}.
\newblock {\url{https://doi.org/10.1007/s10909-018-2078-x}}.

\bibitem[Oliver(1987)]{Oliver1987}
Oliver, J.E., Standard atmosphere.
\newblock In {\em Climatology}; Springer US: Boston, MA,  1987; pp. 801--803.
\newblock {\url{https://doi.org/10.1007/0-387-30749-4_167}}.

\end{thebibliography}

%


\PublishersNote{}
\end{adjustwidth}
\end{document}